\documentclass[amsmath,amssymb,aps,prd,showpacs,nofootinbib]{revtex4}
\usepackage{graphicx}
\usepackage{amsfonts}
\usepackage{amssymb}
\usepackage{amsmath}
\usepackage{bm}
\usepackage{latexsym}

\newcommand{\beq}{\begin{eqnarray}}
\newcommand{\eeq}{\end{eqnarray}}
\newcommand{\be}{\begin{equation}}
\newcommand{\ee}{\end{equation}}

\def\fun#1#2{\lower3.6pt\vbox{\baselineskip0pt\lineskip.9pt
\ialign{$\mathsurround=0pt#1\hfil ##\hfil$\crcr#2\crcr\sim\crcr}}}

\newcommand{{\SD}}{\rm SD}

\newcommand{\vex}{\mbox{\boldmath${x}$}}

\newcommand{\ver}{\mbox{\boldmath${r}$}}

\newcommand{\vep}{\mbox{\boldmath${p}$}}

\newcommand{\veu}{\mbox{\boldmath${u}$}}
\newcommand{\vev}{\mbox{\boldmath${v}$}}

\begin{document}

\title{The leptonic widths of high $\psi$-resonances in unitary
coupled-channel model}

\author{\firstname{A.M.}~\surname{Badalian}}
\email{badalian@itep.ru}
\affiliation{Institute of Theoretical and Experimental Physics,
Moscow, Russia}

\author{\firstname{B.L.G.}~\surname{Bakker}}
\email{b.l.g.bakker@vu.nl}
\affiliation{Department of Physics and Astronomy,
Vrije Universiteit, Amsterdam, The Netherlands}

\date{\today}

\begin{abstract}
The leptonic widths of high $\psi$-resonances are calculated in a
coupled-channel model with unitary inelasticity, where analytical
expressions for the mixing angles between $(n+1)\,^3S_1$ and  $n\,^3D_1$
states and probabilities $Z_i$ of the $c\bar c$ component are
derived. These factors depend on energy (mass) and can be different for
$\psi(4040)$ and $\psi(4160)$. However, our calculations give a small difference
between the mixing angles, $\theta(\psi(4040))= (28^{+1}_{-2})^\circ$
and $\theta(\psi(4160))= (29^{+2}_{-3})^\circ$, and $\sim 10\%$ difference
between the probabilities $Z_1\,(\psi(4040))=0.85^{+0.05}_{-0.02}$ and
$Z_2\,(\psi(4160))=0.79\pm 0.01$. It provides the leptonic widths
$\Gamma_{ee}(\psi(4040))=(1.0\pm 0.1)$~keV, $\Gamma_{ee}(\psi(4160))=
(0.62\pm 0.0.07)$~keV in agreement with experiment; for $\psi(4415)$
$\Gamma_{ee}(\psi(4415))=(0.66\pm 0.06)$~keV is obtained, while for
the missing resonance $\psi(4510)$ we predict its mass, $M(\psi(4500))=(4512\pm 2)$~MeV,
and $\Gamma_{ee}(\psi(4510))=(0.68\pm 0.14)$~keV.
\end{abstract}

\maketitle

\section{Introduction}
\label{sec.1}

The high $\psi$-resonances, $\psi(4040)$, $\psi(4160)$, and
$\psi(4415)$, occur far above the open-charm threshold and their
masses, total widths, and leptonic widths (LWs) are known from the
total cross section of $e^+e^-\rightarrow {\rm hadrons}$ \cite{ref1,ref2,ref3}
and exclusive $e^+e^-$ processes \cite{ref4,ref5,ref6}. The PDG
\cite{ref3} gives their  masses with a good accuracy, better than
10 MeV, however, the discussion on the true values of their leptonic widths
continues \cite{ref6,ref7}.  In particular, four different solutions
of LWs,  which equally well describe the BES data \cite{ref2}, are
presented in Ref.~\cite{ref7}.  Also some parameters, recently
extracted from the Belle data on exclusive $e^+e^-$ processes to
open-charm decay channels \cite{ref6}, differ from those given by
the PDG  \cite{ref3}.  In Table~\ref{tab.01} we summarize the values of the LWs
extracted from different experiments, from which one can see that
there exists a large uncertainty  in the LW of $\psi(4040)$, which
can vary from 0.66 keV to 1.6 keV.

\begin{table}[!htb]
\caption{The leptonic widths of high $\psi$-resonances (in keV), extracted from
different experiments.\label{tab.01}}
\begin{center}
\label{tab.01}\begin{tabular}{|c|c|c|c| }
\hline
    data        & $\Gamma_{ee}(\psi(4040))$ & $\Gamma_{ee}(\psi(4160))$ &
 $\Gamma_{ee}(\psi(4415))$ \\
\hline
 BES \cite{ref2}    &     $0.81\pm 0.20$     & $ 0.50\pm 0.27$     &
   $0.37\pm 0.14$ \\
\hline
 BES  \cite{ref7}  &     0.66 to 1.40     &    0.42 to 1.09    &
    0.45 to 0.77 \\
\hline
 PDG \cite{ref3}    &   $0.86\pm 0.07$    &  $0.48\pm 0.22$     &
  $0.58\pm 0.07$ \\
\hline
 Belle  \cite{ref6}  &  $1.6\pm 0.3$     &   $0.7\pm 0.4$      &
    $1.4\pm 0.3$  \\
\hline
\end{tabular}
\end{center}
\end{table}

The theory of charmonium properties is developing already for
forty years, mostly in different potential models, relativistic
\cite{ref8,ref9,ref10,ref11,ref12} and nonrelativistic
\cite{ref13,ref14,ref15,ref16,ref17}, where calculations of high
excitations are mostly performed in closed-channel approximation,
neglecting open decay channels.  Surprisingly, the predicted masses
appear to be weakly dependent on the model used and mostly agree
with each other and the experimental values, within  $\pm (20-40)$~MeV
(see Table \ref{tab.02}). This result can easily be interpreted.
Consider charmonium in a nonrelativistic model with the Cornell
potential and  then vary the  $c$-quark mass and parameters of the
$Q\bar Q$ potential in a special way. Then even identical spectra
can be obtained \cite{ref13}. However, such freedom in the choice
of parameters does not agree with fundamental ideas about the true
value of the $c$-quark mass and  the $Q\bar Q$ static potential,
and therefore additional physical restrictions on the parameters
must be put using new fundamental results \cite{ref18,ref19}.

In Table \ref{tab.02} we give the masses of high  $n\,^3S_1, m\,^3D_1$
charmonium states, obtained in closed-channel approximation and
using a linear confining potential.

\begin{table}[!htb]
\caption{ The masses of high vector charmonium  states (in MeV) in
relativistic (R) and nonrelativistic (NR) models}
 \begin{center}
\label{tab.02}\begin{tabular}{|c|c|c|c|c|c|c| }
\hline
     State   & NR [11] & R [6]  &  R [8] & NR [14] & R (this paper) & exp. \\
\hline
 $M(3\,^3S_1)$  & 4110  & 4100   &  4095  &  4100    &  4112   & $4039\pm 1$\\
\hline
 $M(2\,^3D_1)$  &  4190  &  4194  & 4191  &  4150    & 4195  & $4191\pm 5$ \\
\hline
 $M(4\,^3S_1)$ &   4460  &   4450   &  4433  &  4445 &  4467  & $4421\pm 4$ \\
\hline
 $M(3\,^3D_1)$ &  -   &    4520    & 4505  &  4525  &   4527  &  absent \\
\hline
\end{tabular}
\end{center}
\end{table}
Here one can see that with exception of  $M(3\,^3S_1)$,  the theoretical values
coincide with the experimental masses of the resonances with $\sim (20-40)$~MeV accuracy
and  therefore one may expect that the mass shifts of the
$\psi$-resonances due to open channels are not large, $\sim(20-50)$~MeV.
Such not so large mass shifts were predicted in the $C^3$ model
\cite{ref15}. Notice that if instead of a linear potential, the
so-called screened confining potential is used \cite{ref20,ref21,ref22},
very large mass shifts, $\sim (100-150)$~MeV, are obtained, e.g.
$M(4\,^3S_1) =4273$~MeV, $M(3\,^3D_1)=4317$~MeV in Ref.~\cite{ref20},
and $M(4\,^3S_1)=4389$~MeV, $M(3\,^3D_1)=4426$~MeV in Ref.~\cite{ref22}.
Specific features of the screened potential will be discussed later.

In the $\psi$-family one resonance, originating  from the $3\,^3D_1$
state, is not observed yet, although its predicted mass,
$M(3\,^3D_1)=(4510\pm 20)$~MeV, lies in the region which was already
studied in different $e^+e^-$ experiments \cite{ref4,ref5,ref6,ref23}.
Here we would like to notice that in exclusive experiments of the
Belle Collaboration \cite{ref6}: $e^+e^-\rightarrow D^+D^{*-}, D^{*+}D^{*-}$,
one can see a wide peak (structure) in the region 4.5 GeV $< \sqrt{s}
< 4.6$ GeV. These data were analyzed, using the $\psi(4040)$,
$\psi(4160)$, and $\psi(4415)$ resonances \cite{ref6}. However, in that
analysis the mass of $\psi(4415)$, $M(\psi(4415))=(4515\pm 18)$~
MeV, appears to be 100 MeV larger than what is found in PDG \cite{ref3} and
other experiments. The interesting point is that this large value
of the mass just coincides with that of the missing $\psi(4510)$,
predicted in different theoretical models (see Table II). To confirm
or exclude the manifestation of $\psi(4510)$ it would be important
to analyze these exclusive reactions, taking into account both
resonances, $\psi(4415)$ with the mass $\sim 4420$ MeV and $\psi(4510)$.
In our calculations the predicted LW of $\psi(4510)$ is not
small, $\Gamma_{ee}(\psi(4510)) \sim 0.6$ keV, if the $3\,^3D_1$
state has rather large admixture of the  $4\,^3S_1$ state \cite{ref12}.

In the present paper we concentrate on the LWs of high  $\psi$-resonances
and relate their parameters to the many-channel picture. Their
values appear to be sensitive to the $c$-quark mass taken and the
parameters of the $Q\bar Q$ static potential. As shown in Ref.~\cite{ref24},
the squared wave functions (w.fs.) at the origin of the $3\,^3S_1$ and  $2\,^3D_1$
states can change several times for different $Q\bar Q$ potentials,
giving very different LWs.  Including the asymptotic-freedom
behaviour of the strong coupling is also important, decreasing the w.f.
at the origin by $\sim 30\%$ \cite{ref25}.

For those reasons we choose here a static potential defined only in
terms of fundamental parameters derived in pQCD \cite{ref18} and
the field correlator method \cite{ref26,ref27}. In particular, we
pay attention to the fact that the new value of the QCD constant
for $n_f=3$, $\Lambda_{\overline{MS}}(n_f=3)=(339\pm 10)$~MeV
\cite{ref18}, is rather large and gives rise to a large QCD vector
constant, $\Lambda_{\rm V}$, defining the strong vector coupling $\alpha_{\rm
V}$, since these constants are interrelated,
$\Lambda_V(n_f=3)=1.4752~\Lambda_{\overline{MS}}=(500\pm 15)$~MeV.  In
our previous analysis a smaller $\Lambda_{\rm V}$ \cite{ref28} was
used, while a larger value of  $\Lambda_{\rm V}=500$~MeV increases the
w.fs. at the origin and LWs of $\psi$ resonances.

\section{Mixing of the $n\,^3S_1$ and $(n-1)\,^3D_1$ states}
\label{sec.2}

The w.fs. at the origin of  pure $n\,^3D_1$ states (defined as
$R_{nD}(0)=\frac{5 R''(0)}{2\sqrt{2} m_{\rm Q}^2}$) are known to
be very small \cite{ref16,ref28} and give rise to small LWs. For
example,  the LW of pure $1\,^3D_1$ is  about seven times smaller than that
of $\psi(3773)$ \cite{ref29}. To explain such a difference a mixing angle
$\theta_1$ between the $2\,^3S_1$ and $1\,^3D_1$ states,  $\theta_1=(11\pm 1)^\circ$,
was extracted from the ratio of their LWs,
$\eta_1=\frac{\Gamma_{ee}(\psi(3773))}{\Gamma_{ee}(\psi(3686))}$
\cite{ref30,ref31}. In the same manner, the mixing angle
$\theta_2=34^\circ$ between the  $3\,^3S_1$ and $2\,^3D_1$ states
was extracted in Ref.~\cite{ref12}, giving $\Gamma_{ee}(4040)=(0.86\pm
0.07)$~keV and $\Gamma_{ee}(\psi(4160))=0.83\pm0.06$~keV  in good
agreement with the old experimental data on LWs from PDG (2006)
\cite{ref32}.  An almost identical value, $\theta_2=37^\circ$, was
obtained in Ref.~\cite{ref30}.  However, now a smaller LW of
$\psi(4160)$ is given by the PDG (2014) \cite{ref3} (see also the
LWs in Table \ref{tab.01}), while the LW of $\psi(4040)$ remains
unchanged,
% Eq. 1
%
\begin{equation}
 \Gamma_{ee}(\psi(4160)) = (0.48\pm 0.22)~{\rm keV}, \quad
 \Gamma_{ee}(4040)) = (0.86\pm 0.07)~{\rm keV}.
\label{eq.1}
\end{equation}
Their ratio also becomes smaller and has a large experimental error,
% Eq. 2
%
\begin{equation}
 \eta_2(\exp.) = \frac{\Gamma_{ee}(\psi(4160))}{\Gamma_{ee}(\psi(4040))}
 = 0.56\pm 0.30,
\label{eq.2}
\end{equation}
i.e.,  $\eta_2$ changes in a wide range, from 0.26 to 0.86, and
therefore is not useful in our analysis.

To extract the mixing angle the resonance w.fs. are usually taken in a
simplified form:
% Eq. 3
%
\begin{equation}
 |\psi(4040) \rangle = |3\,^3S_1\rangle \cos\theta_2 - |2\,^3D_1 \rangle
 \sin\theta_2,\quad
 |\psi(4160) \rangle = |3\,^3S_1\rangle \sin\theta_2 + |2\,^3D_1 \rangle
 \cos\theta_2,
\label{eq.3}
\end{equation}
with equal angles $\theta_2$ in both w.fs. This assumption is not
supported by  results following from  coupled-channel models,
where resonance w.fs. are given by more complicated expressions
\cite{ref33,ref34,ref35} and can schematically be written as
% Eq. 4
%
\begin{eqnarray}
 \varphi(\psi(4040)) &= & \sqrt{Z_1} (\varphi(3S)\cos\theta_2 -
 \varphi(2D)\sin\theta_2)  + \sqrt{1-Z_1}~\varphi_{\rm cont},
\nonumber \\
 \varphi(\psi(4160)) & = & \sqrt{Z_2} (\varphi(3S)\sin\tilde{\theta}_2  +
 \varphi(2D)\cos\tilde{\theta}_2 ) + \sqrt{1-Z_2}~\varphi_{\rm cont},
\label{eq.4}
\end{eqnarray}
and in general contain different mixing angles
$\theta_2=\theta(\psi(4040))$, $\tilde{\theta}_2=\theta(\psi(4160))$
and different probabilities of the $c\bar c$ component, $Z_1=Z(\psi(4040))$
and $Z_2=Z(\psi(4160))$. Besides, in these w.fs. a contribution
from a continuum w.f., $\varphi_{\rm cont}$, is also present.  The
analytical expressions of  $\theta_2$, $\tilde{\theta}_2$, $Z_1$,
and $Z_2$ will be derived in Section \ref{sec.3}, using the
coupled-channel model with unitary inelasticity (CCUI model)  and
taking into account five strong decay channels, $D\bar D$, $D\bar
D^*$, $ D^*\bar D^*$, $D_s\bar D_s$, and $D_s\bar D_s^*$, while here we present
some formulas derived later. First, dynamical calculations give different
mixing angles and  probabilities, because these quantities
are defined at a certain energy, equal to the mass of a given
resonance.  For example,
% Eq. 5
%
\begin{equation}
 Z_1=Z_1(E=M(\psi(4040))), \quad Z_2=Z_2(E=M(\psi(4160))).
\label{eq.5}
\end{equation}
This type of probabilities $Z_{c\bar c}$ was already calculated in the $C^3$
model with the Cornell potential \cite{ref15}, where equal values $Z_{c\bar
c}(\psi(4040))=Z_{c\bar c}(\psi(4160))=0.494(3)$ were obtained.
Surprisingly, in the $C^3$ model the mixing angle  between the $3\,^3S_1$
and $2\,^3D_1$ states was found to be very small, $\theta_2\leq
4^\circ$, much smaller than  the mixing angle, $\sim 16^\circ$,
between the $3\,^3S_1$ and $2\,^3S_1$ states. In our calculations
the mixing angles $\theta_2$ and $\tilde{\theta}_2$ are not small and appear to be
close to each other (see Section \ref{sec.5}).

When the LWs are considered, in the w.f. at the origin a contribution
from the continuum (four-quark or  meson-meson component) can be
neglected, since this contribution is very small \cite{ref36}.
Nevertheless, the influence of decay channels on the w.fs. is kept
through the factors $Z_1$ and $Z_2$ and in general  the resonance w.fs.
at the origin contain four parameters,
% Eq. 6
%
\begin{eqnarray}
 \varphi(\psi(4040),r=0) & = & \sqrt{Z_1} (\varphi(3S) \cos\theta_2 -
 \varphi(2D)\sin\theta_2),
\nonumber \\
 \varphi(\psi(4160),r=0) & = & \sqrt{Z_2}(\varphi(3S) \sin\tilde{\theta}_2 +
 \varphi(2D)\cos\tilde{\theta}_2),
\label{eq.6}
\end{eqnarray}
where the  w.fs. at the origin, $\varphi(3S)$ and $ \varphi(2D)$,
are calculated here using the relativistic string Hamiltonian (RSH) (see
Section \ref{sec.4}). When the factors $Z_i$ are present, then they have to be included to
the standard expression of the LW \cite{ref37},
% Eq.  7
%
\begin{equation}
 G_{ee}(\psi(M_V)) =
 \frac{4 e_c^2\alpha^2}{M_{\rm V}^2} |R(\psi(M_V),r=0)|^2 Z_i \beta_{\rm QCD}.
\label{eq.7}
\end{equation}
Notice that in the RSH the w.f. of the $n\,^3D_1$ state is defined as $R_{\rm
{nD}}(0)=\frac{5 R''_{nD}(0)}{2\sqrt{2} \omega_{\rm Q}^2 }$, where $\omega_{\rm Q}$ is
the quark kinetic energy. In Eq.~(\ref{eq.7}) the QCD radiative correction,
$\beta_{\rm QCD}=1 - \frac{16}{3\pi} \alpha_s(\mu)$ \cite{ref38}, is taken the same for
all vector charmonium states, with numerical value
$\beta_{\rm {QCD}}=0.60$; it corresponds to $\alpha_s(n_f=4,\mu)=0.235$ at the scale
$\mu\sim 4$~GeV, if $\Lambda_{\overline{MS}}(n_f=4)=(296\pm 10)$~MeV is taken
from pQCD \cite{ref18}). This factor $\beta_{\rm {QCD}}$ is cancelled in the ratio of the LWs, but
for $\psi(4040)$ and $\psi(4160)$ this ratio cannot be used because of the large
experimental error in $\eta_2$, see Eq.~(\ref{eq.2}).

\section{The mixing angles and probabilities $Z_i$ in the CCUI model}
\label{sec.3}

Here we use the CCUI model \cite{ref33,ref34}, where two sectors are
considered: one refers to the charmonium conventional states and
another to the heavy-light meson sector.  For stationary states, like
$3S$ and $2D$, one can use the Green's function in energy representation,
% Eq. 8
%
\begin{equation}
 G^{(0)}_{Q\bar Q} \left(1,2;\,E\right) =\,\sum_{n_1}
 \frac{\Psi^{(n_1)}_{Q\bar{Q}} (1)\,
 \Psi^{\dag(n_1)}_{Q\bar{Q}}(2)}{E_{n_1}-E}=\,\frac{1}{H_0-E}
\label{eq.8}
\end{equation}
where in the Green's function the superscript (0)  refers to the
bare case, when  the heavy-light sector is switched off. The w.f.
$\Psi_{Q\bar Q}^{n_1}$ and $E_{n_1}$ are the eigenfunctions
and eigenvalues of the relativistic string Hamiltonian (RSH)
\cite{ref39,ref40} (see the next Section).  In the heavy-light sector we use the Green's
function of the pair $(Q\bar q)(q\bar Q)$ and neglect the interaction
of the two (color singlet) heavy-light mesons. Then in the c.m.
system one can write the Green's function as
% Eq. 9
%
\begin{equation}
 G^{(0)}_{Qq\,\bar q\bar Q} \left(1\bar 1 |2\bar 2;\, E\right) =
 \sum_{n_2, n_3} \int
 \frac{\Psi_{n_2\,n_3} (1,\bar 1)\, \Psi_{n_2\,n_3}^\dag(2,\bar 2)}
 {E_{n_2n_3} (\vep)-E}\,\,d\Gamma(\vep),
\label{eq.9}
\end{equation}
where $d\Gamma(\vep)$ is the phase space factor.

To take into account transitions from the $Q\bar Q$ state to the
sector of heavy-light mesons (strong decays) the Lagrangian of
$\,^3P_0$ type is used,
% Eq. 10
%
\begin{equation}
 \mathcal{L}_{sd} = \int \bar \psi_q\, M_\omega\, \psi_q\, d^{\,4}
 x ,\quad M_\omega =const,
\label{eq.10}
\end{equation}
where $M_{\omega}=const.\approx 0.8$~GeV and $\psi_q$  are relativistic
w.fs. of a light quark in the field of a heavy antiquark $\bar Q$
\cite{ref33}. It is important that the w.fs., entering  the Green's
functions and the RSH, are considered in the c.m. system, where the time
coordinates of all particles are the same. Therefore the vertex
$\mathcal{L}_{sd}$ occurs between instantaneous w.fs. of the $Q\bar
Q$ system on one side and the product of the $Q\bar q$ and $q\bar Q$ w.fs.
on the other side, thus defining an overlap integral $J_{123}$:
% Eq. 11
%
\begin{equation}
 J_{123} \equiv \frac{1}{\sqrt{N_c}}\int\bar y_{123} \Psi^{\dagger}_{Q\bar Q}\,
 M_\omega\, \psi_{Q\bar q}\, \psi_{q\bar  Q} \,d\tau.
\label{eq.11}
\end{equation}
Here the factor $\bar y_{123}$ is determined by  the spin and total
angular momentum  of the considered system; its explicit expressions
for charmonium states $n\,^3S_1,~m\,^3D_1$ are given in
Refs.~\cite{ref33,ref34}. The matrix element $J_{123}$ is reduced
to the overlap integral, with $\ver$  proportional to $\veu - \vev$
\cite{ref33}:
% Eq. 12
%
\begin{equation}
 J_{n_1n_2n_3}(\vep) = \frac{M_\omega}{\sqrt{ N_c}} \int \bar y_{123}^{\rm Rel}\,
 \Psi^{(n_1)}_{Q\bar Q} (\veu-\vev)\, e^{i\,\vep \cdot \ver}
 \psi^{(n_2)}_{Q\bar{q}}
 (\veu-\vex)\, \psi^{(n_3)}_{\bar{Q}q} (\vex-\vev)\, d^{\,3} \vex\, d^{\,3}
 (\veu-\vev),
\label{eq.12}
\end{equation}
which defines the self-energy contributions to the mass  of a $Q\bar
Q$ meson, appearing due to heavy-light mesons in the intermediate
states:
% Eq. 13
%
\begin{equation}
 w_{nm} (E) = \int \frac{d^{\,3} \vep}{(2\pi)^3} \sum_{n_2n_3}
 \frac{J_{nn_2n_3}
(\vep)\, J^{\dagger}_{mn_2n_3} (\vep)}{E-E_{n_2n_3}(\vep)}.
\label{eq.13}
\end{equation}
Then the total Green's function (in the $Q\bar Q$ sector)  can be
written as a sum over bound states:
% Eq. 14
%
\begin{equation}
 G^{(I)}_{Q\bar Q} (1,2;\,E) = \sum_n \frac{\Psi^{(n)}_{Q\bar Q} (1)\,
 \Psi^{\dagger(n)}_{Q\bar Q} (2)}{E_n-E}-\sum_{n,m} \frac{\Psi^{(n)}_{Q\bar Q}
 (1)\, w_{nm} (E)\,  \Psi^{\dagger(m)}_{Q\bar Q} (2)}{(E_n-E)(E_m-E)} + \dots
\label{eq.14}
\end{equation}
and the solutions of the equation,
% Eq. 15
%
\begin{equation}
 \det \big(E-\hat E^0-\hat w\big) =0,
\label{eq.15}
\end{equation}
define a new spectrum, namely, the masses $E_{R_1}$ and $E_{R_2}$ of two
resonances in the two-channel case. In the case we consider here, the
index 1 refers to the resonance $\psi(4040)$) and index 2 to the resonance $\psi(4160)$.
In Eq.~(\ref{eq.15}), $\hat E^0$ is a diagonal matrix, $\delta_{nm}E_m^0$,
while the matrix elements  $w_{ik}(E)$ determine the mixing
angle between the $3\,^3S_1$  and $2\,^3D_1$  states and the mass
shifts,  $w_{11}=w_{\rm {SS}}, ~w_{22}=w_{\rm {DD}}$. We will also
use the notations: $E_{11}^*= E_1^0 + w_{11}(E)$ and $E_{22}^*=E_2^0
+ w_{22}(E) $ using the initial masses $E_n^0$ of the $3\,^3S_1$
and $2\,^3D_1$ states, calculated in closed-channel approximation.

The matrix elements $w_{ik}(E)$, dependent on the energy, are taken at the
energy equal to the resonance mass: $E=E_{\rm {R_1}}$ for $\psi(4040)$
and $E=E_{\rm {R_2}}$ for $\psi(4160)$. If the non-diagonal matrix elements are
small, then in first approximation the masses and widths of the
resonances are $ E_{\rm {R1}} = E_1^0 + {\rm {Re}}\, w_{11}(E_1^0)$
and $ E_{\rm {R2}} = E_2^0 +{\rm {Re}}\,w_{22}(E_2^0)$, $\Gamma_{\rm
{R1}}= 2~{\rm {Im}}\, w_{11}(E_1^0)$ and  $\Gamma_{\rm {R2}}=
2~{\rm{Im}}\,w_{22}(E_2^0)$.  However, for high $\psi$-resonances
the values of the non-diagonal m.es.  $w_{ik}$ are not small, being
only about two times smaller than the  mass differences.
The matrix Eq.~(\ref{eq.15}) can be diagonalized, introducing a
unitary matrix $\hat U$:
% Eq. 16
%
\begin{equation}
 \big((E-\hat E -\hat w)^{-1}\big)_{nm} = U^\dagger_{n\lambda} (E)
 \frac{1}{E-E_\lambda}\, U_{\lambda m} (E),
\label{eq.16}
\end{equation}
defining a set of resonance w.fs. $\Phi_{\lambda}$.
Then the Green's function acquires the new form:
% Eq. 17
%
\begin{equation}
 G_{Q\bar Q} ^{(I)} = \sum_\lambda \Phi_\lambda \frac{1}{E_\lambda-E}\,
 \Phi^\dagger_\lambda , \quad \Phi_\lambda = \sum_n \Psi^{(n)}_{Q\bar Q}\,
 U^\dagger_{n\lambda}(E),
\label{eq.17}
\end{equation}
i.e., the  w.fs. $\Phi_{\lambda}~(\lambda=1,2)$ become new orthogonal
states, comprising all effects of the mixture between bound states
owing to the decay channels. The same procedure can be applied to the
states above the decay thresholds, if one neglects the widths of
those states. Here we use just this approximation.

Using Eqs.~(\ref{eq.16},\ref{eq.17}) one can  express the new w.fs.
$\Phi_{\lambda}$ via the w.fs. $\Psi_n$, taking into account the
following relations,
% Eq. 18
%
\begin{equation}
 (E- \hat {E}^0 -\hat w)^{-1} = \frac{1}{\det(E - \hat {E}^0- \hat w)}
  \left(
 \begin{array}{cc}
 E - E_1^0 - w_{11} & w_{21}            \\
 w_{12}             & E - E_2^0 - w_{22}
 \end{array}
 \right).
\label{eq.18}
\end{equation}
With the notations  $E_2^*= E_2^0 + w_{22}(E)$, $E_1^*=E_1^0 +
w_{11}(E)$, and $E_{{\rm R}_i}=E_{\lambda_i}$,   we rewrite $\det(E-\hat{E}^0
-\hat w)$ as $\det(E - \hat{E}^*) =(E - E_{\lambda_1})(E -E_{\lambda_2})=
(E - E_1^*) (E - E_2^*) - w_{12}(E) w_{21}(E)$, which defines the
masses of the two resonances. Now we assume that $E_{\lambda_1} <
E_{\lambda_2}$. Then the mass of the resonance with the smaller mass is
% Eq. 19
%
\begin{equation}
 E_{\lambda_1} = \frac{1}{2}(E_1^* + E_2^*) -
 \frac{1}{2}\sqrt{(E_2^*- E_1^*)^2  + 4 w_{12} w_{21}}
\label{eq.19}
\end{equation}
where  all matrix elements $w_{12}(E)$,  $w_{21}(E)$, $ w_{11}(E)$, and
$w_{22}(E)$  inside $E_n^*$ are taken at the point $E=E_{\lambda_1}$.
The mass of the higher resonance, $E_{\lambda_2}$ is obtained from
the equation,
% Eq. 20
%
\begin{equation}
 E_{\lambda_2} = \frac{1}{2}( E_1^* + E_2^*) +
 \frac{1}{2}\sqrt{(E_2^* - E_1^*)^2  +4 w_{12} w_{21}},
\label{eq.20}
\end{equation}
with all matrix elements $w_{ik}(E)$, taken at the point $E=E_{\lambda_2}$.

To find the explicit expressions of the matrix elements of the unitary matrix
$U_{ik}$ we assume that the  imaginary parts of $w_{ik}$ are small
and can be omitted. Then the inverse matrix in Eq.~(\ref{eq.18})
is written as
% Eq. 21
%
\begin{equation}
 (E - \hat {E}^*)_{11}^{-1} = \sum_{\lambda=1,2} U_{1\lambda}^{\dagger}(E)
 \frac{1}{E - E_{\lambda}} U_{\lambda_1}(E)=  \frac{1}{E-E_{\lambda_1}}  +
 \frac{E_{\lambda_2} - E_2^*}{E_{\lambda_1} - E_{\lambda_2}}
 \left(\frac{1}{E - E_{\lambda_1}} - \frac{1}{E-E_{\lambda_2}}\right).
\label{eq.21}
\end{equation}
Therefore the products of the matrix elements are
% Eq. 22
%
\begin{equation}
 U_{11}^\dagger U_{11} = 1 + \frac{E_{\lambda_2} -E_2^*}
 {E_{\lambda_1} - E_{\lambda_2}}, \quad
 U_{12}^\dagger U_{21} = - \frac{E_{\lambda_2} - E_2^*}
 {E_{\lambda_1} - E_{\lambda_2}}
\label{eq.22}
\end{equation}
Notice that these matrix elements satisfy the property of unitarity: $U_{11}^\dagger
U_{11} + U_{12}^\dagger U_{21}=1$. In the same way we find the
product of the other matrix elements,
% Eq. 23
%
\begin{equation}
 U_{21}^\dagger U_{12} = \frac{E_{\lambda_1} - E_1^*}
{E_{\lambda_1} - E_{\lambda_2}},\quad
 U_{22}^\dagger U_{22}=1 - \frac{E_{\lambda_1} - E_1^*}
{E_{\lambda_1} - E_{\lambda_2}},
\label{eq.23}
\end{equation}
which satisfy the condition $U_{21}^\dagger U_{12} + U_{22}^\dagger
U_{22}=1$. From the relations (\ref{eq.22}) and (\ref{eq.23}) the
diagonal matrix elements are found to be
% Eq 24
%
\begin{eqnarray}
 U_{11} & = &\sqrt{1 + \frac{E_{\lambda_2} - E_2^*}
 {E_{\lambda_1} - E_{\lambda_2}}},
\nonumber \\
 U_{22} & = & \sqrt{1 - \frac{E_{\lambda_1} - E_1^*}
 {E_{\lambda_1} - E_{\lambda_2}}}.
\label{eq.24}
\end{eqnarray}
The non-diagonal matrix elements are given by
% Eq. 25
%
\begin{equation}
 U_{12}=\frac{w_{21}}{(E_{\lambda_1} - E_{\lambda_2}) U_{11}}, \quad
 U_{12}^\dagger = \frac{w_{21}}{(E_{\lambda_1} - E_{\lambda_2}) U_{22}},
\label{eq.25}
\end{equation}
and satisfy the condition: $U_{21}^\dagger U_{11} + U_{22}^\dagger
U_{21} =0$. From the obtained expressions the w.f. $\Phi_1$ of the
lower resonance with the mass $E_{\lambda_1}$   and the w.f. $\Phi_2$
of the upper resonance with the mass $E_{\lambda_2}$
($E_{\lambda_1} < E_{\lambda_2}$), can be written as
% Eq. 26
%
\begin{equation}
 \Phi_1(\psi(4040)) = U_{11} \left(\Psi_1 - \Psi_2
 \frac{w_{12}}{E_2^*- E_{\lambda_1}}\right) =
 \sqrt{Z_1} \left( \Psi_1 \cos\theta_1 + \Psi_2 \sin\theta_1\right),
\label{eq.26}
\end{equation}
where we have taken into account that $w_{12}= - |w_{12}|$ and introduced
% Eq. 27
%
\begin{equation}
 \tan\theta_1 = \frac{|w_{12}|}{E_2^* - E_{\lambda_1}},\quad
 \sin\theta_1= \frac{|w_{12}|}{\sqrt{(E_2^* -E_{\lambda_1})^2 + w_{12}^2}}.
\label{eq.27}
\end{equation}
The probability $Z_1$, given by
% Eq. 28
%
\begin{equation}
 Z_1= \frac{(E_2^* - E_{\lambda_1})^2 +
 w_{12}^2}{(E_{\lambda_2}- E_{\lambda_1})(E_2^* - E_{\lambda_1})},
\label{eq.28}
\end{equation}
determines the weight of the state 1 ($3\,^3S_1 $) in the resonance
w.f. of $ (\psi(4040))$ (where all matrix elements are taken at $E=E_{\lambda_1}$).
Its value is close to unity if the transition matrix element  $|w_{12}|$ is
small, $w_{12}^2 \ll (E_2^* - E_{\lambda_1})^2$. However, in a
realistic situation where $|w_{12}|\sim (30-50)$~MeV, it can be of the same
order as the mass difference, $E_2^*-E_{\lambda_1}\sim 100$~MeV.

For the w.f. of the upper resonance an expression similar to
Eq.~(\ref{eq.27}) applies,
% Eq. 29
%
\begin{equation}
 \Phi_2(\psi(4160)) = \sqrt{Z_2}
 (- \Psi_1 \sin\tilde{\theta}_2 + \Psi_2 \cos\tilde{\theta}_2),
\label{eq.29}
\end{equation}
with
% Eq. 30
%
\begin{equation}
 Z_2 =\frac{(E_{\lambda_2} - E_1^*)^2 + w_{21}^2}
 {(E_{\lambda_2} -E_{\lambda_1})   (E_{\lambda_2}  - E_1^*)}
\label{eq.30}
\end{equation}
and
% Eq. 31
%
\begin{equation}
 \sin\tilde{\theta}_2=\frac{|w_{21}(E)|}{\sqrt{(E_{\lambda_2} - E_1^*)^2 +
 w_{21}^2}}.
\label{eq.31}
\end{equation}
In Eqs.~(\ref{eq.29}-\ref{eq.31}) all matrix elements are taken
at the energy $E=E_{\lambda_2}$.

Thus for mixed $(n+1)\,^3S_1$ and $n\,^3D_1$ states  analytical
expressions were derived, which allow to calculate the mixing angles
and the probabilities in the  w.fs. of the $\psi(4040)$ and $\psi(4160)$
resonances and to understand dynamical effects, produced by the five
decay channels $D\bar D$, $D\bar D^*$, $D^*\bar D^*$, $D_sD_s$, and
$D_sD_s^*$. These relations and the condition $Z_i<1.0$ establish
important correlations between the different mass shifts.

Also we would like to notice that in our calculations the sign of
the w.f. at the origin is taken with the factor $(-1)^{n+1} =
(-1)^{n_r+l}~(n_r=0, 1,...)$ as in simple harmonic oscillator (SHO)
functions.

\section{The static potential}
\label{sec.4}

From the analytical expressions Eqs.~(\ref{eq.27}-\ref{eq.31}) one
can see that the parameters of the resonances explicitly depend
on the bare masses $E_1^0$ and $E_2^0$, defined  by the static
potential $V_0(r)$. In our approach this potential contains only fundamental quantities,
established in pQCD \cite{ref18} and the field correlator method
\cite{ref26,ref27}, and owing to the so-called Casimir scaling
$V_0(r)$ has to be the sum of the confining and gluon-exchange terms
\cite{ref27}.  In the confining term the string tension
$\sigma_0=(0.18\pm 0.02)$~GeV$^2$ is fixed by the slope of the leading
Regge trajectory of light mesons, while the asymptotic freedom
behaviour of the vector coupling is defined by the QCD vector
constant $\Lambda_V(n_f=3)$ in full agreement with the value
$\Lambda_{\overline{MS}}(n_f=3)=(339\pm 10)$~MeV from pQCD \cite{ref18}, since
they are interrelated.  Namely, $\Lambda_{\rm V}(n_f=3) =1.4753
\Lambda_{\overline{MS}}(n_f=3)=(500\pm 15)$~MeV \cite{ref19}.

At small momenta in two-loop vector coupling $\alpha_{\rm V}(q^2)$
we take the value of the infrared regulator $M_{\rm B}$, which
enters the logarithm $\ln\left(\frac{q^2+M_{\rm B}^2}{\Lambda^2_{\rm
V}}\right)$, from Ref.~\cite{ref41}, to be  $M_{\rm B} = \sqrt{2
\pi\sigma_0}$, which is not an extra parameter, since it is expressed
through the same string tension $\sigma_0$ as occurs in the leading
Regge trajectory of light mesons.

It is of interest to notice that for $\Lambda_{\rm V}=(480\pm 20)
$~MeV and $M_{\rm B}=(1.10\pm 0.05)$~GeV the critical (asymptotic)
vector coupling,  $\alpha_{\rm crit}=\alpha(q^2=0)= 0.60\pm
0.04$ has the value close to $\alpha_{\rm crit}=0.60$ in the Godfrey-Isgur
model \cite{ref9}.

In closed-channel approximation we use the RSH \cite{ref39}, where
in the kinetic term  the pole $c$-quark mass, $m_c({\rm
pole})=(1.440\pm 0.015)$~GeV corresponds to the conventional current
quark mass, $\overline{m}_c =(1.267\pm 0.011)$~GeV \cite{ref3}.  Then for
the $Q\bar Q$ potential,
% Eq. 32
%
\begin{equation}
   V_0(r) = \sigma_0 r  - \frac{4\alpha_{\rm V}(r)}{3r},
\label{eq.32}
\end{equation}
the centroid masses $M_{\rm {cog}}(nl)$ coincide with the eigenvalues of the
spinless Salpeter equation (SSE):
% Eq. 33
%
\begin{equation}
  \left(2 {\sqrt{\vep^2 + m_Q^2} +V_0(r)} \right) \psi(r) = M_{\rm {cog}}(nl) \psi(r).
\label{eq.33}
\end{equation}
Thus the charmonium spectrum is defined without fitting parameters.
In Eq.~(\ref{eq.32}) the two-loop coupling $\alpha_{\rm V}(r)$ in
coordinate space is expressed via the two-loop vector coupling in
momentum space,
% Eq. 34
%
\begin{equation}
 \alpha_{\rm V}(r) = \frac{2}{\pi} \int \limits_0^\infty {\rm d}q
 \frac{\sin(qr)}{q}\, \alpha_{\rm V}(q^2),
\label{eq.34}
\end{equation}
and its properties were studied in detail in Ref.~\cite{ref19}.
Here we take the following set of the parameters,
% Eq. 35
%
\begin{eqnarray}
 \sigma_0 & = & 0.18~{\rm GeV}^2,~~\Lambda_V(n_f=3) = 500 ~{\rm MeV}
 \;{\rm or}\;
 \Lambda_{\overline{MS}}(n_f=3) = 339~{\rm MeV}, \quad
\nonumber \\
 M_B & = & 1.15~ {\rm GeV}, \quad\alpha_{\rm crit}=0.635,
 \quad m_c = 1.440~{\rm GeV}.
\label{eq.35}
\end{eqnarray}
In  the CCUI model the masses of the resonances and the mass shifts
are determined via the ``unperturbed" masses. They include spin
corrections and in Section~\ref{sec.3} were denoted as $E_1^0=M(3\,^3S_1)=
M_{\rm {cog}}(3S) + 1/4 \Delta_{\rm {hf}}$, with $\Delta_{\rm
{hf}}(3S)$ being the hyperfine shift of the $3\,^3S_1$ state, and
$E_2^0=M(2\,^3D_1)= M_{\rm {cog}}(2D) - \Delta_{\rm {fs}}$, where
$\Delta_{\rm {fs}}$ is a shift of the $2\,^3D_1$ state due to the
fine-structure interaction.

In $V_0(r)$, Eq.~(\ref{eq.32}), we use a linear confining potential
(without flattening or screening effects), in order to escape double
counting in the coupled-channel calculations. About the screened potential,
% Eq. 36
%
\begin{equation}
 V_{\rm scr}=\lambda r \frac{1-\exp(-\mu r)}{\mu r},
\label{eq.36}
\end{equation}
it is worth to notice that it is going to a constant at not so large
distances, $R\sim 2$~fm: $V_{\rm scr}(r)\rightarrow
const.=\frac{\lambda}{\mu}=2.145$~GeV \cite{ref20}. Such an asymptotic
behavior violates the boundary conditions of the relativistic
SSE as well as the Schr\"{o}dinger equation, and
makes the gluon-exchange potential dominant even at large distances.
Besides, in a high charmonium state with the mass, $M (nl) \geq
(2m_c + 2.145) \sim (5.0  - 5.2)$ GeV, its constituents, a quark and
an antiquark, are not confined but can be liberated.

In Table~\ref{tab.03} we give the w.fs. at the origin, calculated
in closed-channel approximation with the static potential $V_0(r)$,
which are needed for further coupled-channel analysis, and also
the LWs of pure $n\,^3S_1$, $m\,^3D_1$ states, taking the factor $\beta_{\rm
{QCD}}=0.60$ for all states. The unperturbed masses $E_i^0$ are
also given.

The w.fs.  $R_{nS}(0)$ are calculated here in two ways, because the
original form of the string Hamiltonian \cite{ref19,ref28,ref39,ref40},
% Eq. (37)
%
\begin{equation}
 H_{\rm str}= \omega + m_Q^2/\omega + \vep^2/\omega  + V_0(r),
\label{eq.36a}
\end{equation}
has to be supplemented by the extremum condition, which can be of
two different kinds.  In the first case the condition $\partial
H_{\rm str}/\partial\omega=0$ is used, which allows to reduce  $H_{\rm str}$
to the Hamiltonian $H_0$, present in the SSE,  Eq.~(\ref{eq.33}).
This equation, when solved numerically \cite{ref10}, is very convenient,
since it gives simultaneously the whole meson spectrum, the quark
kinetic energies $\omega_{nl}$, all matrix elements, etc. However,
for the SSE the radial w.fs. $R_{nS}(0)$ have an unpleasant feature
-- they diverge near the origin for any potential $V_0(r)$ with
Coulomb-type term. Therefore, a regularization of $R_{nS}(r)$ at small
distances, which can produce additional fitting parameters, is needed.
Here we use a procedure which allows to escape the introduction of new parameters
doing regularization.

For the SSE at small $r$ the derivatives $R_{nS}'(r)$ increase, starting
to grow at a critical distance $r_{nS} \sim 0.07$ fm. On the contrary,
the eigenfunctions of the $H_{\rm str}$ are regular in so-called einbein approximation
\cite{ref28,ref39,ref40} (as well as those of the Schr\"{o}dinger equation)
and their derivatives decrease for small $r$, approaching zero at $r=0$. In the einbein approximation
to $H_{\rm str}$, Eq.~(\ref{eq.36a}), the extremum condition is put
on the meson mass $M_{nl}: \partial M_{nl}(\omega_{nl})/ \partial \omega_{nl}=0$,
where the mass is $ M_{nl}= \omega_{nl} + m_Q^2/\omega_{nl} +E_{nl}$ and this extremum
condition defines the quark kinetic energy $\omega_{nl}$ as
%Eq.38
%
\begin{equation}
 \omega_{nl}^2= m_Q^2 + \omega_{nl}^2 \partial E_{nl}/\partial \omega_{nl},
\label{eq.36b}
\end{equation}
while the eigenvalues are the solutions of the equation,
%Eq. (39)
%
\begin{equation}
 (\vep^2/ \omega +V_0(r)) \psi_{nl}(r) = E_{nl} \psi_{nl}(r).
\label{eq.36c}
\end{equation}
Thus in the einbein approximation one needs to solve in a consistent way
two equations, Eqs.~(\ref{eq.36b} and \ref{eq.36c}). The numerical calculations of the eigenvalues
has an accuracy of $\sim 1$ MeV. Moreover, the w.fs. at the origin $R_{nS}^{EA}(0)$ can also
be defined with the use of the relation \cite{ref8}:
%Eq. (40)
%
\begin{equation}
 |R_{nS}^{EA}(0|^2 = \omega_{nS} \langle dV_0/dr \rangle_{nS} =
 \omega_{nS} ( \sigma_0 + 4/3 \langle \alpha(r)/r^2 \rangle _{nS}  -
 4/3 \langle \alpha'(r)/  r \rangle_{nS} ) ,
\label{eq.36d}
\end{equation}
where all matrix elements are calculated with great accuracy. Notice that the relation
(\ref{eq.36d}) contains the quark kinetic energy $\omega_{nl}$, which depends on the quantum numbers,
while in the nonrelativistic case, instead, the quark mass $m_Q$ enters for all states.

At small $r$ the derivatives $R'_{nS}(EA,r)$ are very small, approaching zero,
and their values can be used for regularization of the SSE w.fs.. Using this procedure
one obtains the regularized w.fs. $R_{nS}^{\rm reg}(0)$ of the SSE, which values
appear to be very close to $R_{nS}^{\rm EA}(0)$, calculated in einbein approximation.
Due to this fact and to escape additional uncertainties coming from the regularization,
we use here $R_{nS}^{\rm EA}(0)$ as the w.f. at the origin of SSE, taking the proper
value of the quark kinetic energy $\omega_{nS}$.

\begin{table}[!htb]
\caption{The w.fs. $R_{\rm {nl}}(0)$ (in GeV$^{3/2}$), the masses $M(nl)$
 (in MeV), and the leptonic widths (in keV) of pure $n\,^3S_1$ and
 $m\,^3D_1$ charmonium states for the static potential $V_0(r)$ and
 $m_c=1.44$~GeV, $\beta_{\rm {QCD}}=0.60$}
\begin{center}
\label{tab.03}\begin{tabular}{|c|c|c|c|c|c|}
\hline
   state  & $R_{\rm {nl}}(0)$&$\Gamma_{ee}(th.)$ &$\Gamma_{ee}(exp.)$\cite{ref3} &
 $ M_{\rm V}(nl)$ &
   experiment\\
\hline
   $1\,^3S_1$  &  0.961 &  5.47    &   $5.55\pm 0.14$     &  3093   &  $3096.90\pm 0.01$\\
   $2\,^3S_1$  & -0.801 &  2.68   &   $2.34\pm 0.04 $     &  3689   & $3686.10\pm 0.03$ \\
   $1\,^3D_1$  &  0.096 &  0.037 &   $0.262\pm 0.018$  &  3800   & $3773.13\pm 0.35$ \\
   $3\,^3S_1$  &  0.752 &  1.97   &   $0.86\pm 0.07$      &  4112   & $4039\pm 1$  \\
   $2\,^3D_1$  & -0.146 & 0.069 &   $0.48\pm 0.22 $      &  4195   & $4191\pm 5$  \\
   $4\,^3S_1$  & -0.738 &  1.58  &    $0.58\pm 0.07$      &  4467   & $4421\pm 4$ \\
   $3\,^3D_1$  & 0.170 &  0.080 &    -                 &  4527   &    -   \\
\hline
\end{tabular}
\end{center}
\end{table}

As seen from Table~\ref{tab.03}, good agreement with experiment is
obtained only for the LW of $J/\psi$. For $\psi(3686)$ the LW is
$15\%$ larger, while the LWs of the other $n\,^3S_1$ states are about two times larger than
the experimental numbers.  On the other hand, the LWs of pure
$n\,^3D_1~(n=1,2)$ states are about seven times smaller than
the experimental values of $\Gamma_{ee}(\psi(3773))$ and $\Gamma_{ee}(\psi(4190))$.
It is of interest to notice that for the gluon-exchange potential with
large vector constant, $\Lambda_{\rm V}=500$~MeV, the w.fs. at the
origin are relatively large and owing to that,
$\Gamma_{ee}(J/\psi)=5.47$~keV is found to be in good agreement
with the experimental value.

For the set of parameters, Eq.~(\ref{eq.35}), the unperturbed masses are calculated
numerically with accuracy $\sim 1$~MeV (see also Table~\ref{tab.02}),
% Eq. 41
%
\begin{equation}
 E_1^0=M(3\,^3S_1)=4112~{\rm MeV}; ~~ E_2^0=M(2\,^3D_1)=4195~{\rm MeV}.
\label{eq.37}
\end{equation}
and for the  $4\,^3S_1$ and $3\,^3D_1$ states we find
% Eq. (42)
%
\begin{equation}
 E_1^0(4\,^4S_1)=M(4\,^3S_1)=4467~{\rm MeV},\quad
 E_2^0(3\,^3D_1)=M(3\,^3D_1)= 4527~{\rm MeV}.
\label{eq.38}
\end{equation}
These values are used further in the CCUI model.

Now we compare the  parameters we have chosen , Eq.~(\ref{eq.35}), with those
from Ref.~\cite{ref28}, where in the vector coupling $\alpha_V$ a
smaller $\Lambda_V(n_f=4) =360$~MeV is used. Notice, that this value
of $\Lambda_V$ corresponds to $\Lambda_{\overline{MS}}(n_f=4)=253$~MeV,
which is $\sim 17$\% smaller
than $\Lambda_{\overline{MS}}(n_f=4))=(296\pm 10)$~MeV, accepted now in pQCD
\cite{ref3,ref18}. Therefore, this QCD constant has to be considered as a fitting parameter.
Also a significant difference takes place between the value of the
QCD factor, $\beta_{QCD}=0.72$ in Ref.~\cite{ref28} and $\beta_{QCD}=0.60$,
used in the present analysis for all states.

In Table~\ref{tab.04}  we compare the LWs, calculated here and in Ref.~\cite{ref28},
and give also the values of the radial w.fs. at the origin for low-lying
charmonium states. For $\psi(3686)$ and $\psi(3773)$ the $2S-1D$ mixing angle,
$\theta=11.5^\circ$, is taken here, while in Ref.~\cite{ref28}  $\theta
=11^\circ$;  also for $\Lambda_V=360$~MeV and $M_B=1.0$~GeV,
taken in \cite{ref28}, the critical (frozen) $\alpha_{\rm crit}=0.547$
is smaller than in our present case, where $\alpha_{\rm crit}=0.635$.

Nevertheless, in Ref.~\cite{ref28} the values of $R_{nS}(0)$  appeared
to be  only $\sim (5-7)$\% smaller than in our case.
For the $1S$, $2S,$ and $1D$ states their values (in GeV${}^{3/2}$) are
0.905, 0.767, 0.094 in Ref.~\cite{ref28} and 0.962, 0.801, 0.096 in our calculations.
Therefore, if the same $\beta_{QCD}$ is taken, then the LWs in
Ref.~\cite{ref28} would be $10-15$\% smaller than in our case.
For that reason, to fit the experimental LWs a larger $\beta_{QCD}=0.72$ is taken
in Ref.~\cite{ref28}. This choice of $\beta_{QCD}=1- \frac{16}{3\pi}\alpha_s(\mu)
 = 0.72$ cannot be considered as a good one, since it corresponds to
a very small $\alpha_s(\mu)(n_f=4)=0.165$, or to a very large scale $\mu > 8.0$~
GeV.  In our calculations $\beta_{QCD}=0.60$ is smaller and corresponds to
$\alpha_s(\mu)(n_f=4) =0.235$ with a reasonable value of the scale
$\mu \sim (3.7 - 4.0)$ GeV, which is in agreement with pQCD.

In our calculations of the LWs, presented in Table~\ref{tab.04}, their values are
defined by Eq.~(\ref{eq.7}), while in \cite{ref28}  the LWs contain an additional
relativistic factor $\xi_R= \frac{m^2 +\omega_{nl}^2 +\vep^2/3}{2\omega_{nl}
^2}$, originating from the vector decay constant expression \cite{ref42}.
Therefore, for a comparison it is convenient to divide the LWs
of $J/\psi$, $\psi(3686)$, and $\psi(3773)$ from Ref.~\cite{ref28} by the values
of $\xi_R$, equal to 0.929, 0.910, 0.910, respectively.
.

\begin{table}[!htb]
\caption{Comparison of the radial w.fs. at the origin (in GeV${}^{3/2}$, first three rows)
and the leptonic widths (in keV) for low-lying charmonium states.
The mixing angle $\theta=11^\circ$ in Ref.~\cite{ref28} and
$\theta=11.5^\circ$ in the present paper}
\begin{center}
\label{tab.04}\begin{tabular}{|c|c|c| c|}
\hline
 State                    & \cite{ref28} & this paper & experiment, \cite{ref3}\\
 \hline
 $J/\psi$                 & 0.905                 & 0.961 & \\
 $\psi(3686)$             & 0.735                 & 0.764 & \\
 $\psi(3773)$             & 0.238                 & 0.246 & \\
 $\Gamma_{ee}(J/\psi)$    & 5.82                  & 5.47  & $5.55\pm 0.14$ Ref.~\cite{ref14} \\
 $\Gamma_{ee}(\psi(3686)$ & 2.71                  & 2.44  & $2.34\pm 0.04$ Ref.~\cite{ref4} \\
 $\Gamma_{ee}(\psi(3773)$ & 0.27                  & 0.242  & $0.262\pm 0.018$ Ref.~\cite{ref18}\\
\hline
\end{tabular}
\end{center}
\end{table}

A comparison of the LWs with the experimental data \cite{ref3} shows
that for our set of parameters the LWs of low-lying states are
obtained in good agreement with experiment.

Notice that the existing uncertainty in the value of the QCD constant
$\Lambda_V(n_f=3)=(500\pm 15)$~MeV does not change $R_{nS}(0)$ by more
than 1\%.  There is also an uncertainty in the
value of the infrared regulator, $M_B=(1.07\pm 0.08)$~GeV, and here we fix
$M_B=1.15$ GeV according to the analysis of the bottomonium spectrum in Ref.~\cite{ref19}.

It is important to stress that the vector coupling in
coordinate space,  Eq.~(34),  is taken with $n_f=3$, while in
momentum space the regions with different $q^2$ are described by
the strong coupling with different numbers of flavours $n_f$. In coordinate space
the situation is different, because the ``exact " (or combined) coupling $\alpha_C(r)$,
defined by Eq.~(\ref{eq.34}), coincides with $\alpha_V(n_f=3, r)$ for all distances
with the exception of very small $r < 0.06$~fm \cite{ref19}, so that the use of $\alpha_V(n_f=3, r)$
in the whole region provides high accuracy in the charmonium masses and w.fs..
On the contrary, a choice of the vector couping with $n_f=4$ and
$\Lambda_V(n_f=4)=360$ MeV in coordinate space \cite{ref28} has no fundamental grounds.

\section{Results}
\label{sec.5}

In the CCUI, the  calculation of the matrix elements $w_{ik}$, defined by the overlap integral
$J_{nn_2n_3}$, Eq.~(\ref{eq.13}), is the most important part of the
numerical calculations. In this overlap integral we approximate
the exact w.fs., expanding them in a series of simple harmonic
oscillator (SHO) functions and take five terms for the charmonium
w.fs. and one SHO function for heavy-light mesons.  All parameters
of these SHO functions are given in Ref.~\cite{ref33}. The accuracy
of the numerical calculations is estimated to be $\sim 10\%$.

Since the matrix elements $w_{ik}(E)$  depend on the energy, they differ at
the points $E=E_{\lambda_1}=M(\psi(4040))=4056$~MeV and
$E=E_{\lambda_2}=M(\psi(4160))=4190$~MeV. In Table \ref{tab.04a}
we give the values of $w_{ik}$ with the errors arising from numerical
calculations. Notice that the best agreement with the experimental LWs
is reached not for the central values, but for the maximal values of $|w_{ik}|$.
We also introduce new notations: $w_{11}=w_{\rm{SS}}, w_{12}=w_{\rm {SD}},
w_{22}= w_{\rm {DD}}$.

\begin{table}[!htb]
\caption{The matrix elements $w_{ik}$ (in MeV) which take into account
the five decay channels,
$D\bar D,~D\bar D^*, ~D^*\bar D^*,  ~D_sD_s$,  and $D_sD_s^*$.}
\begin{center}
\label{tab.04a}\begin{tabular}{|c|c|c| }
\hline
   $ E_{\lambda} $ &   4056   &  4190   \\
\hline
 $ w_{\rm {SS}} $ & $-32\pm 5$     &  $-2\pm 1$     \\
 $ w_{\rm {SD}} $ &  $- 46\pm 3$   &  $-45\pm 6$    \\
 $ w_{\rm {DD}} $ &  $-50\pm 5$   & $ -30\pm 6$   \\
\hline
\end{tabular}
\end{center}
\end{table}
Then from Eqs.~(\ref{eq.19}, \ref{eq.20}) the masses of the resonances are as follows,
% Eq. 43
%
\begin{equation}
 M(\psi(4040))=(4056\pm 8)~{\rm MeV}, \quad M(\psi(4160))=(4190^{+3}_{-1})~{\rm MeV},
\label{eq.39}
\end{equation}
i.e., the central value of $M(\psi(4040))$ is obtained to be 56 MeV lower
than the initial mass of  the $3\,^3S_1$ state. The situation is different for
the higher solution of Eq.~(\ref{eq.20}), when this mass almost coincides with the mass of the
$2\,^3D_1$ state; it happens because this mass decreases due to the self-energy
shift $w_{DD}$ but increases owing to the non-diagonal matrix element $w_{SD}$.

Using  Eqs.~(\ref{eq.27}) and (\ref{eq.31}) and $w_{ik}$ from Table~\ref{tab.04a}, one
obtains that the mixing angles $\theta_2$ and $\tilde{\theta}_2$ have rather close values,
% Eq. 44
%
\begin{equation}
 \theta_2(\psi(4040))=(28^{+1}_{-2})^\circ; ~~ \tilde{\theta}_2(\psi(4160))=(29.5^{+2.0}_{-3.0})^\circ.
\label{eq.40}
\end{equation}
Although the difference between these angles is small, the use of
them gives rise to LWs in better agreement with the experimental
values. Using these mixing angles and the w.fs. at the origin from
Table~\ref{tab.03}, one obtains the following radial w.fs. at the origin
of the resonances: $R(\psi(4040),r=0)=(0.593^{+0.006}_{-0.008})$~GeV$^{3/2}$
and $R(\psi(4160),r=0)= (- 0.496^{+0.023}_{-0.031})$~GeV$^{3/2}$.
Then the probabilities $Z_i$ and the LWs are
% Eq. 45
%
\begin{equation}
 Z_1(\psi(4040)=0.86^{+0.04}_{-0.03}, \quad
 \Gamma_{ee}(\psi(4040))=Z_1 (1.21^{+0.02}_{-0.03})~{\rm keV}=(1.04^{+0.07}_{-0.06})~{\rm keV}.
\label{eq.41}
\end{equation}
Here the central value of the LW is $\sim 10\%$ larger than

the upper limit of the experimental value $\Gamma_{ee}(\psi(4040))=(0.86\pm 0.07)$~{\rm keV},
given by the PDG \cite{ref3}, but smaller than the LW, obtained in the analysis of the BES data \cite{ref7}
and the Belle data \cite{ref6} (see Table ~\ref{tab.01}).  For the $\psi(4160)$ resonance we have found
% Eq. 46
%
\begin{equation}
 Z_2=(0.79\pm 0.01), \quad \Gamma_{ee}=Z_2 (0.80^{+0.07}_{-0.10})~{\rm keV}=(0.62\pm +0.07)~{\rm keV}.
\label{eq.42}
\end{equation}

We have also checked the sensitivity of our results to the choices
of the $c$-quark mass and of $\Lambda_{\rm V}$, varying them in a very
narrow range, since the pole $c$-quark mass and $\Lambda_{\rm V}$ are known
with $\pm 20$~MeV accuracy in pQCD. To describe the masses of low-lying charmonium states
with a good accuracy (with the smaller $c$-quark mass, $m_c=1.425$~GeV), one needs
to use the value of $\Lambda_{\rm V}=465$~MeV, which is smaller than that accepted in pQCD,
$\Lambda_V(n_f=3)=(500\pm 15)$~MeV. Also in this case the matrix elements  $|w_{ik}|$ are a bit
smaller, than those in Table~\ref{tab.04a}, giving the smaller $\theta_2=23^\circ$, $M(\psi(4040))=4034$~MeV, and $Z_1=0.97$, so that the calculated value $\Gamma_{ee}(\psi(4040))=Z_1 1.30=1.26$~keV is larger
compared to the value in Eq.~(\ref{eq.41}). For the higher resonance the mass $M(\psi(4160))=4160$~MeV
is also 30 MeV smaller than in Eq.~(\ref{eq.37}), while $Z_2=0.78$ and the mixing angle,
$\tilde{\theta}_2=30^\circ$, is not changed, giving the same value of
$\Gamma_{ee}(\psi(4160))=Z_2 0.81~{\rm keV} =0.63$~keV, as in Eq.~(\ref{eq.42}).

For the set of the parameters from Eq.~(\ref{eq.35}) our results are summarized in Table~\ref{tab.05}.

\begin{table}[!htb]
\caption{The mixing angles $\theta_2$ and $ \tilde{\theta}_2$,
 the probabilities $Z_1,$ and $Z_2$, and the leptonic widths (in keV)
 of the $\psi(4040)$ and $\psi(4160)$ resonances}
\begin{center}
\label{tab.05}\begin{tabular}{|c|c|c|c|c|c|c|}
\hline
 state     & $E^0$ &$ M_{\rm R}$ & $\theta_i$ & $ Z_i$ & $\Gamma_{ee}$& exp.\\
\hline
 $\psi(4040)$ & $4112\pm 1$ & $4056\pm 8$ & $(28^{+1}_{-2})^\circ$ & $0.86^{+0.04}_{-0.03}$ & $1.0\pm 0.1$ & $0.86\pm 0.07$\\
 $\psi(4160)$ & $4195\pm 1$ & $4190^{+3}_{-1}$ & $(29^{+2}_{-3})^\circ$ & $0.79\pm 0.01$ & $0.62\pm 0.07$ & $0.48\pm 0.22$\\
\hline
\end{tabular}
\end{center}
\end{table}
From Table~\ref{tab.05} one can see that for the  chosen $Q\bar Q$ potential, Eq.~(\ref{eq.36})
with the set of the parameters Eq.~(\ref{eq.35}) and the pole mass $m_c({\rm pole})=1.44$~GeV, the calculated
LWs are found to be in better agreement with the experimental values of  BES \cite{ref2} and
PDG \cite{ref3}, than in case of the smaller $m_c=1.425$~GeV, where the central values are
% Eq. 47
%
\begin{equation}
 \Gamma_{ee}(\psi(4040))=Z_1 1.30 =1.26~{\rm keV}, \quad
 \Gamma_{ee}(\psi(4160))=Z_2 0.78 =0.63~{\rm keV}.
\label{eq.43}
\end{equation}
In this case  $\Gamma_{ee}(\psi(4040))\sim 1.3$~keV appears to be close to the LW value
extracted from the Belle \cite{ref6} and BES experimental data \cite{ref7}. Thus the
existing disagreement between the experimental data on the LWs
of $\psi(4040)$ does not allow to fix the values of the
$c$-quark mass and the QCD  constant $\Lambda_{\rm V}(n_f=3)$ with high accuracy.

\begin{table}[!htb]
\caption{The leptonic widths (in keV) of the $\psi(4415)$ and $\psi(4500)$ resonances}
\begin{center}
\label{tab.06}\begin{tabular}{|c|c|c|c|c|c|c|}
\hline
 state             & $E^0$ & $M_{\rm R}$       & $\theta_i$              & $Z_i$                             & $\Gamma_{ee}$            & exp.\\
\hline
 $\psi(4415)$ & 4467  & $4421^{+7}_{-8}$ & $(33\pm 4)^\circ$ & $0.83^{+0.06}_{-0.03}$ & $0.66^{+0.06}_{-0.05}$ & $0.58\pm 0.07$ \\
 $\psi(4510)$ & 4527  & $4512\pm 2$        & $(34\pm 3)^\circ$ & $0.85^{+0.08}_{-0.06}$ & $0.68\pm 0.14$           & -  \\
\hline
\end{tabular}
\end{center}
\end{table}
Calculation of the LW of $\psi(4415)$ with a good accuracy is a more
difficult task, first, because $\psi(4510)$ is not observed yet,
and secondly, because the mass difference,  $M(4510)-M(4420)=90$~MeV
is rather small and therefore all matrix elements $w_{ik}$ and  $Z_i$ have
to be determined with great accuracy. Our calculations (with accuracy
$\sim10\%$) give the following $w_{ik}$ at the point $E=4421$~MeV:
$w_{11}=(-23\pm 2)$~MeV, $w_{12}=(-35\pm 5)$~MeV, $w_{22}=(-53\pm 5)$~MeV, and
% Eq. 48
%
\begin{eqnarray}
 \theta_3(\psi(4415))& =& (33\pm 4)^\circ, \quad \tilde{\theta}_3=(34\pm 3)^\circ,
 \quad Z_3(\psi(4415)) = 0.83^{+ 0.06}_{-0.03}, \quad   Z_4(\psi(4510)=0.85^{+0.08}_{-0.06},
\nonumber \\
 R(\psi(4415),r=0) & = & (- 0.52\pm 0.04)~{\rm GeV}^{3/2},\quad
 R(\psi(4510),r=0)=(0.53\pm 0.04)~{\rm GeV}^{3/2},
\label{eq.44}
\end{eqnarray}
to obtain the values
\begin{equation}
 \Gamma_{ee}(\psi(4415))=Z_3 (0.79\pm 0.02)= (0.66^{+0.06}_{-0.05})~{\rm keV}\quad
 \Gamma_{ee}(\psi(4510))=Z_4 (0.80\pm 0.10)~{\rm keV}=(0.68\pm 0.14)~{\rm keV},
\label{eq.45}
\end{equation}
i.e. for $\psi(4415)$ and $\psi(4510)$ the LWs are almost equal.

Thus we conclude that the $4S-3D$ mixing, occurring via open $DD$,
$DD^*$, and  $D^*D^*$ channels, may not be small, with the mixing angle $\sim
33^\circ$, which is very close to the mixing angle $\theta_{ph}=34.5^\circ$, used
in  a phenomenological approach \cite{ref12}. Then for the missing state $\psi(4510)$
the LW $(0.68\pm 0.10)$~keV is obtained, which is almost  equal to that of $\psi(4415)$.

Our calculations were done in a simplified two-channel model, where the coupling to meson-meson
channels and also to $c\bar c$ vector channels, like $2\,^3S_1$ and $1\,^3D_1$,
is not taken into account. Such many
channel considerations are very complex within the analytical CCUI model.
However, our dynamical calculations
allow us to define mixing angles, which are usually taken as fitting
parameters, and also the probabilities of the $c\bar c$ components
$Z_i$, which partly suppress the values of the LWs. Although these factors
depend on energy, for the resonances
$\psi(4040)$ and $\psi(4160)$ the mixing angles appear to be almost equal,
while the probabilities $Z_i$ in $\psi(4040)$ and $\psi(4160)$) can differ by
up to $15\%$.

In the CCUI model the analysis of LWs is of special importance, because
the w.fs. at the origin do not depend on the admixture of a meson-meson (a multiquark)
component, which nevertheless could decrease the probabilities $Z_1$
and $Z_2$ (or $Z_3$ and $Z_4$).

\section{Summary}
\label{sec.6}

We have used a coupled-channel model with unitary inelasticity to
describe the mixing of $n\,^3S_1$ and $(n-1)\,^3D_1$ states, which
occurs due to transitions to decay channels. In this model analytical
expressions for the resonance masses, mixing angles, and probabilities,
needed for understanding the physical picture, are obtained. For
the $\psi$-resonances the following masses are calculated:
$M(\psi(4040)) =(4056\pm 8)$~MeV, $M(\psi(4160))=(4190^{+3}_{-1})$~MeV,
$M(\psi(4415))=(4421^{+7}_{-8})$~MeV, and $M(\psi(4510))=(4512\pm 2)$~MeV.

At present there is no  consensus about the precise value of the LWs of
$\psi(4040)$, $\psi(4160)$ and in different analyses of the experimental
data on inclusive and exclusive $e^+e^-$ processes two possibilities
are presented: a relatively large LW $\Gamma_{ee}(\psi(4040) \approx
1.2$~keV in Refs.~\cite{ref6,ref7} and a smaller value of
$\Gamma_{ee}(\psi(4040))=0.86(7)$~keV in Refs.~\cite{ref2,ref3}. In
our model $\psi(4040)$ and $\psi(4160)$ have almost equal mixing angles
(they coincide within errors): $\theta(\psi(4040))\cong \theta(\psi(4160))
\cong (29\pm 2)^\circ$, nevertheless, a small difference
between the mixing angles provides better agreement with the experimental data.

An important suppression of the LWs is possible due to the probabilities $Z_i$,
which for $\psi(4040)$ and $\psi(4160)$ differ by only $\sim 10\%$. However, 

these values are sensitive to the transition matrix elements and can
vary in a wide range, from 0.72 to 0.93 within accuracy of calculations. For the 
LWs our calculations give  $\Gamma_{ee}(\psi(4040))=(1.0\pm 0.1)$~keV, and
$\Gamma_{ee}(\psi(4160))=(0.62\pm 0.07)$~keV.

We have also considered the mixing between $4\,^3S_1$ and $3\,^3D_1$ via decay channels.
It appears to be sufficiently strong, producing a rather large mixing angle:
$\theta\sim (33\pm 4)^\circ$, so that the LWs of $\psi(4415)$ and the missing resonance
$\psi(4510)$ have almost equal LWs:  $\Gamma_{ee}(\psi(4415))=(0.66\pm 0.06)$~keV and
$\Gamma_{ee}(\psi(4510))=(0.68\pm 0.14)$~keV.

\begin{acknowledgments}
The authors are very grateful to Yu.~A.~Simonov for important discussions and
clarifications of some aspects of the CCUI model. The authors are grateful to V.~D.~Orlovsky
for important help in numerical calculations.
\end{acknowledgments}

\end{document}